\documentclass[sigconf, authors]{acmart}

\usepackage[]{algorithm2e} 

\usepackage{algorithmic} 
\usepackage[nolist]{acronym}
\usepackage{tabularx}
\usepackage{rotating}
\usepackage{multirow}
\usepackage{listings}
\usepackage{xcolor}
\usepackage{float}
\usepackage{lipsum}    
\usepackage{graphicx}  
\usepackage{subcaption}
\usepackage{todonotes}
\usepackage{fancybox}
\usepackage{booktabs}

\usepackage{array}

\usepackage{tikz}
\usepackage{pgfplots}
\usepackage{pgfplotstable}
\pgfplotsset{compat=1.11,
    /pgfplots/ybar legend/.style={
    /pgfplots/legend image code/.code={%
       \draw[##1,/tikz/.cd,yshift=-0.25em]
        (0cm,0cm) rectangle (3pt,0.8em);},
   },
}

\usepackage{amsmath}

\newcolumntype{C}[1]{>{\centering\arraybackslash}p{#1}}

\usetikzlibrary{arrows}

\def\checkmark{\tikz\fill[scale=0.4](0,.35) -- (.25,0) -- (1,.7) -- (.25,.15) -- cycle;} 

\definecolor{codegreen}{rgb}{0,0.6,0}
\definecolor{codegray}{rgb}{0.5,0.5,0.5}
\definecolor{codepurple}{rgb}{0.58,0,0.82}
\definecolor{backcolour}{rgb}{0.95,0.95,0.92}

\lstdefinestyle{mystyle}{
    commentstyle=\color{codegreen},
    keywordstyle=\color{magenta},
    numberstyle=\tiny\color{codegray},
    stringstyle=\color{codepurple},
    basicstyle=\ttfamily\footnotesize,
    breakatwhitespace=false,         
    breaklines=true,                                     
    keepspaces=true,                 
    numbers=left,                    
    numbersep=5pt,                  
    showspaces=false,                
    showstringspaces=false,
    showtabs=false,                  
    tabsize=1
}

\copyrightyear{2023} 
\acmYear{2023} 
\setcopyright{acmlicensed}\acmConference[ACSAC '23]{Annual Computer Security Applications Conference}{December 4--8, 2023}{Austin, TX, USA}
\acmBooktitle{Annual Computer Security Applications Conference (ACSAC '23), December 4--8, 2023, Austin, TX, USA}
\acmPrice{15.00}
\acmDOI{10.1145/3627106.3627138}
\acmISBN{979-8-4007-0886-2/23/12}

\begin{document}

\title{On the Feasibility of Cross-Language Detection of Malicious Packages in npm and PyPI}

\author{Piergiorgio Ladisa}
\affiliation{%
  \institution{SAP Security Research}
  \streetaddress{805, avenue du Dr Maurice Donat}
  \city{Mougins}
  \country{France}
  \postcode{06250}
}
\affiliation{%
  \institution{University of Rennes 1/INRIA/IRISA}
  \streetaddress{263 avenue du Général Leclerc}
  \city{Rennes}
  \country{France}
  \postcode{35 042}
}
\email{piergiorgio.ladisa@sap.com}
\email{piergiorgio.ladisa@irisa.fr}

\author{Serena Elisa Ponta}
\affiliation{%
  \institution{SAP Security Research}
  \streetaddress{805, avenue du Dr Maurice Donat}
  \city{Mougins}
  \country{France}
  \postcode{06250}
}
\email{serena.ponta@sap.com}

\author{Nicola Ronzoni}
\affiliation{%
  \institution{SAP Security Research}
  \streetaddress{805, avenue du Dr Maurice Donat}
  \city{Mougins}
  \country{France}
  \postcode{06250}
}
\email{nicola.ronzoni@sap.com}

\author{Matias Martinez}
\affiliation{%
  \institution{Universitat Politècnica de Catalunya - Barcelona Tech}
  \streetaddress{Carrer de Jordi Girona, 31}
  \city{Barcelona}
  \country{Spain}
  \postcode{08034}
}
\email{matias.martinez@upc.edu}

\author{Olivier Barais}
\affiliation{%
  \institution{University of Rennes 1/INRIA/IRISA}
  \streetaddress{263 avenue du Général Leclerc}
  \city{Rennes}
  \country{France}
  \postcode{35 042}
}
\email{olivier.barais@irisa.fr}

\renewcommand{\shortauthors}{Piergiorgio Ladisa et al.}

\begin{abstract}

  Current software supply chains heavily rely on open-source packages hosted in public repositories.
  Given the popularity of ecosystems like npm and PyPI, malicious users started to spread malware by publishing open-source packages containing malicious code. 
  
  Recent works apply machine learning techniques to detect malicious packages in the npm ecosystem.
  However, the scarcity of samples poses a challenge to the application of machine learning techniques in other ecosystems. 
  Despite the differences between JavaScript and Python, the open-source software supply chain attacks targeting such languages show noticeable similarities (e.g., use of installation scripts, obfuscated strings, URLs).
  

  In this paper, we present a novel approach that involves a set of language-independent features and the training of models capable of detecting malicious packages in npm and PyPI by capturing their commonalities. This methodology allows us to train models on a diverse dataset encompassing multiple languages, thereby overcoming the challenge of limited sample availability. 
  We evaluate the models both in a controlled experiment (where labels of data are known) and in the wild by scanning newly uploaded packages for both npm and PyPI for 10 days.  
  
  We find that our approach successfully detects malicious packages for both npm and PyPI.
  Over an analysis of 31,292 packages, we reported 58 previously unknown malicious packages (38 for npm and 20 for PyPI), which were consequently removed from the respective repositories.


  


\end{abstract}



\maketitle

\section{Introduction}

From large companies to independent developers, the current way of producing software is characterized by a large consumption of open-source packages.
Given the popularity of some programming languages, package repositories and package managers have been created for specific ecosystems to ease the consumption of \ac{OSS} for downstream users.
Package repositories (e.g., PyPI, npm) are public databases that can be queried to retrieve packages implementing certain tasks. On the client side, package managers automatically resolve and install the required packages as well as their dependencies.

While these mechanisms facilitate software modularization and increase the speed of implementation, malicious users started to exploit these mechanisms as an easy way to spread malwares on a large scale. 
The attack surface of the \ac{OSS} supply chain is large and attackers have multiple attack vectors at their disposal to conduct \textit{open-source software supply chain attacks}~\cite{ladisa2022taxonomy}. 
Given that most companies (both private and public) use open-source~\cite{githubOctoverse2022}, enhancing the security of the software supply chain has become a priority in the interest of both the community and nations security~\cite{sonatypeAnnualState, whitehouseExecutiveOrder, enisaThreatLandscape2022}.

One way to counter \ac{OSS} supply chain attacks is to detect the presence of malicious behavior in packages consumed from package repositories. 
Recent works propose \ac{ML} techniques for vetting packages in npm~\cite{sejfia2022practical,ohm2022feasibility}. 
The primary challenge in applying \ac{ML} approaches to detect malicious packages lies in the availability of labeled data. Currently, the widely used dataset for malicious packages is \ac{BKC}~\cite{dasfreakBackstabbersKnife}, which comprises 102 distinct malicious packages for JavaScript and 92 for Python (as of October 2022).



Though every language has its own characteristics (e.g., a particular list of sensitive APIs), we observe that malicious packages from the Python and JavaScript ecosystems present common patterns. As examples, the code snippets responsible for the malicious behavior are usually characterized by the presence of obfuscated strings and the usage of installation scripts. 

In this preliminary work, we evaluate the feasibility of using a machine-learning approach to capture similarities of malicious behaviors across different languages, specifically Python and JavaScript. This entails constructing a \textit{cross-language} classifier using a dataset featuring samples from both languages, as well as developing two separate \textit{mono-language} classifiers trained on datasets comprising samples from either JavaScript or Python.


We set out to answer the following research questions:

\textbf{RQ1:} Which cross-language and mono-language models demonstrate the best trade-off between precision and recall when detecting malicious packages in the case of JavaScript and Python languages? 

\textbf{RQ2:} How do the models identified in RQ1 perform in the detection of potentially malicious packages in the wild? 

Our work's technical contributions in addressing these questions are as follows.

Firstly, we propose the use of 141 language-independent features, derived from a combination of previous works~\cite{sejfia2022practical,ohm2022feasibility} as well as expert knowledge.

Secondly, we evaluate tree-based learning algorithms, specifically \ac{DT}, \ac{RF}, and \ac{XGBoost}, to train three classifiers for distinct datasets: one comprising solely JavaScript packages, another containing only Python packages, and a third containing packages from both languages. 
Out of the 9 models generated, we discovered that both the cross-language and mono-language models based on \ac{XGBoost} performed best in a controlled experiment. 

Thirdly, we conduct a real-world assessment of these models by analyzing packages uploaded daily to npm and PyPI over a 10-day period.  
This experiment revealed that the cross-language model outperformed the respective mono-language models.

Lastly, throughout our analysis involving 31,292 packages, we successfully identified 58 previously unidentified malicious packages across npm and PyPI. We promptly reported these malicious discoveries to the respective repositories, which confirmed our findings. Moreover, we enhanced \ac{BKC}\footnote{Commit \texttt{6d7d140168f80c048fc1622a4788b1f5c17af2d0}} by uploading these newly identified malicious packages.



The paper is organized as follows.
Section~\ref{sec:background} provides an overview of the problem of Open Source Software (OSS) supply chain attacks and highlights the key distinctions between Python and JavaScript.
Section~\ref{sec:methodology} presents the datasets, the learning algorithms employed for training the models to classify malicious packages, and the set of features used.  
Section~\ref{sec:evaluation} answers the research questions. 
Section~\ref{sec:discussion} examines the findings and noteworthy aspects of the npm and PyPI ecosystems observed during the real-world experiment.
Section~\ref{sec:resp-disclosure} delves into the processes involved in responsibly reporting the identified malicious packages.
Section~\ref{sec:threats-validity} outlines potential threats to the validity of the study.
Section~\ref{sec:relatedworks} presents related works within the domain of machine-learning approaches for detecting malicious packages.
Section~\ref{sec:conclusion} summarizes the key conclusions drawn from the study and provides insights into future prospects.

\section{Open-Source Software Supply Chain Attacks}\label{sec:background}

\ac{OSS} supply chain attacks involve the introduction of malicious code into open-source components, which serves as a method for propagating malware among downstream consumers. Various attack vectors can be employed to carry out such attacks~\cite{ladisa2022taxonomy}. Broadly speaking, attackers may choose to: \textit{develop and advertise a malicious package from scratch}, \textit{create name confusion with legitimate package}, or \textit{subvert a legitimate package} (i.e., injecting the malicious code in the source code, during the build, or within the package repository).

Regardless of the attack vector used, attackers aim to have their malicious package(s) hosted on popular distribution platforms such as npm and PyPI registries. Downstream users typically acquire OSS by downloading it from these platforms using dependency management systems like pip or the npm CLI. After victims download a malicious package, the malicious code contained within it can be triggered at various stages of the package lifecycle, including installation, runtime, or during testing~\cite{ohm2020backstabbers}. 

In 2020, Ohm et al.~\cite{ohm2020backstabbers} observed that in most cases (i.e., 56\%) attackers inject malicious code in a manner that is triggered during the installation phase.
One way to achieve this is by leveraging the capability of package managers to automatically execute custom \textit{installation scripts} included within the downloaded packages. These scripts enable the specification of actions to be executed during the installation process. The designated entry point for these scripts is referred to as the \textit{installation hook}.

In JavaScript, the \textit{package.json} file provides installation hooks through the \textit{scripts} property. This property is a dictionary where the keys represent lifecycle phases (e.g., \textit{pre-install}), indicating when the scripts in the corresponding values should be executed~\cite{npmjsPackagejsonDocs}.

In Python, the \textit{setup.py} script is executed during the installation of source distributions (\textit{sdist}). Attackers have the opportunity to inject malicious code anywhere within this file, allowing them to execute commands at installation time~\cite{pythonIntroductionDistutils}.


Attackers may also inject malicious code into other executable scripts within the package, aiming for its execution during runtime or tests whenever the manipulated functionality is triggered.

Depending on the specific attack vector employed by attackers, different safeguards can be implemented to partially or fully mitigate such attacks~\cite{ladisa2022taxonomy}. One effective protection measure involves detecting malicious packages prior to their consumption. 
The academic community has already begun proposing methods for identifying malicious packages~\cite{ohm2022feasibility,10.1145/3407023.3409183,DBLP:journals/corr/abs-2011-02235,sejfia2022practical,garrett2019detecting,ladisajavamalware,duan2021measuring}. A prominent challenge~\cite{filiol2006open} for all detection approaches is the wide variety of evasion techniques that can be employed by malware. 
Therefore, it is crucial to enrich datasets of malicious packages (e.g., \ac{BKC}) such that new strategies can be devised to detect malware categories akin to those already encountered or anticipate emerging types of malware.

\section{Models Training}\label{sec:methodology}

In this section, we outline the training process for both the mono-language and cross-language models, which are designed to classify malicious packages. First, we describe the datasets we constructed, comprising benign and malicious packages in JavaScript and Python. 
Then, we delve into the algorithms and optimizations employed to obtain the models. Finally, we present the collection of language-independent features extracted from the packages.

\subsection{Datasets}\label{sec:datasets}

We build a labeled dataset of benign and malicious packages. Given that benign packages significantly outnumber malicious ones in real package repositories, we create an imbalanced dataset to reflect this reality. As the exact ratio of malicious to benign packages remains uncertain, we adopt the 90-10 ratio between benign and malicious samples, as recommended in previous studies~\cite{shabtai2009detection, moskovitch2008unknown}.

\begin{table}[!hbtp]

  \centering
  \begin{tabular}{lrrrr}

    \textbf{Language} &  \textbf{Total \#}  & \textbf{Filter by }   & \textbf{Filter by} & \textbf{Filter by} \\
  & \textbf{of samples} & \textbf{version} &  \textbf{campaign} &  \textbf{duplicates}\\  \cmidrule(lr){1-5} 
  
  JavaScript& 2071 & 1505 & 1408 & 102\\
  Python  & 273 & 225 & 133  & 92  \\

\end{tabular}
  \caption{Number of malicious samples remaining after applying each filtering step.} \label{tab:bkc-filtering}
  \end{table}

\textbf{Malicious samples.} We use \ac{BKC}~\cite{dasfreakBackstabbersKnife}, which is a collection of malicious \ac{OSS} packages found in popular package repositories and voluntarily contributed by the community. At the time of writing (October 2022, commit \texttt{c2d9691}) \ac{BKC} contains a total of 3,161 packages for different ecosystems: 2,071 JavaScript, 273 Python, 813 Ruby, and 4 Java packages. Since our focus is on JavaScript and Python, we restrict to packages for such ecosystems. It is noteworthy that a significant portion of the packages is associated with malware campaigns (i.e., packages having different names but containing the same malicious behavior~\cite{ohm2020backstabbers}). Consequently, it is possible to encounter multiple duplicates within the \ac{BKC} dataset.
To avoid bias we remove duplicates from the dataset. We consider as duplicates packages that \emph{(i)} have multiple versions, \emph{(ii)} are marked as part of a campaign in \ac{BKC}, and \emph{(iii)} have the same values for the considered features (cf. Section~\ref{sec:feature-set}).
In the first case, we consider only the latest version.
For packages belonging to the same campaign or having the same values for the features, we take only one sample. Table~\ref{tab:bkc-filtering} shows the number of packages remaining after each filtering step. Finally, we get 102 malicious packages for npm and 92 for PyPI.

Regarding malicious behaviors, it is worth mentioning that our approach is limited to those present in \ac{BKC}, i.e., reverse shell, dropper, data exfiltration, \ac{DoS}, and financial gain~\cite{ohm2020backstabbers}. Therefore, additional behaviors (e.g., phishing campaigns~\cite{checkmarx140kNuGet}) are out of our scope.

\textbf{Benign samples.} 
Since no ground truth for benign packages exists, we build our dataset as follows.

Assuming that popular projects in npm and PyPI  are free from malicious code~\cite{ohm2022feasibility}, we aim at collecting these projects and use the popularity information provided by \texttt{libraries.io}\footnote{\url{https://libraries.io/}} through the SourceRank score\footnote{\url{https://docs.libraries.io/overview.html\#sourcerank}}.  
Since \texttt{libraries.io} offers APIs to search for packages by category (e.g., machine learning, math, UI) but not by popularity, we adopted the following approach to gather popular packages. First, we compiled a list of 150 categories from the ones suggested by the \texttt{libraries.io} UI for the most popular packages. Second, we searched for these keywords and sorted the results in descending order of popularity. Third, we selected the top-\textit{n} projects for each ecosystem to maintain the desired 90-10 ratio (i.e., 828 for Python and 918 for JavaScript). Finally, we download the packages from the respective package repositories (i.e., npm for JavaScript and PyPI for Python).  

\textbf{Mono-language and Cross-Language Datasets.} Once we have collected both malicious and benign packages, we proceed to construct various types of datasets.
The two mono-language datasets contain packages in a single programming language: one for JavaScript and one for Python. The cross-language dataset consists of the union of the two mono-language datasets. Table~\ref{tab:dataset} shows the number of packages contained in each dataset and related stratification in benign and malicious. 

\begin{table}[!hbtp]

  \centering
  \begin{tabular}{llll}

    \textbf{Type} &  \textbf{Programming}  & \textbf{Malicious}   & \textbf{Benign} \\
  & \textbf{Language(s)} & \textbf{samples} &  \textbf{samples} \\  \cmidrule(lr){1-4} 
  
  Mono-language  & JavaScript &102 & 918 \\
  Mono-language  & Python&92  & 828  \\
  Cross-language & JavaScript+Python   &194 & 1640   \\

\end{tabular}
  \caption{Composition of both mono-language and cross-language datasets.}\label{tab:dataset}
  \end{table}

  \subsection{Algorithms Selection and Tuning}\label{sec:meth-classification}

  The main objective of our work is to explore the feasibility of cross-language detection of malicious packages.
  As mentioned earlier, malicious packages are assumed to be significantly fewer than benign ones. Therefore, we seek supervised classification algorithms that are well-suited for imbalanced datasets, requiring no preprocessing of the dataset itself. Furthermore, to gain more insights into the performances of the models, we look for algorithms that provide explainable predictions.
  Finally, due to the number of features considered, we only consider learning algorithms that handle high-dimensional data.

  Algorithms that meet these criteria and that showed the best performances in recent works are \acl{DT} (DT)~\cite{sejfia2022practical,ohm2022feasibility} and \acl{RF} (RF)~\cite{ohm2022feasibility}. In the context of tree-based algorithms, we also consider the \ac{XGBoost} (XGB) algorithm~\cite{chen2016xgboost}, knowing that boosting algorithms are well-suited for imbalanced datasets~\cite{friedman2001greedy}.
  We train the aforementioned classifiers using \textit{scikit-learn}\footnote{\url{https://scikit-learn.org}}.
  
  To fine-tune the selected algorithms for the classification problem we use the \ac{BO}~\cite{turner2021bayesian} in combination with 5-fold cross-validation to find the best combination of hyperparameters leading to the highest precision. We choose precision as the objective function to reduce the number of false positives, i.e., the number of samples that the security analyst has to manually review to confirm the classification. The value of the precision considered for the optimization problem is the average value obtained after the 5-fold cross-validation.
  
  For the \ac{DT} classifier, the hyperparameters optimized through the \ac{BO} are the maximum depth of the tree, maximum number of features, split criterion (i.e., information gain, Gini index, log-loss), the minimum number of observations within the leaves, and minimum number of samples required to split an internal node. 
  For the \ac{RF} classifier, in addition to the ones for \ac{DT}, we also optimize the number of estimators (i.e., decision trees) and the maximum number of samples to train each tree.
  Finally, for the \ac{XGBoost} classifier, the hyperparameters to be optimized are the maximum depth of the tree, number of estimators, subsample ratio of features to construct each tree, learning rate, minimum split loss, and minimum sum of hessian needed in a child node of the tree.

\subsection{Features} \label{sec:feature-set}

\begin{table*}[!hbtp]

  \centering
  \begin{tabular}{lp{10.5cm}p{4.5cm}}

    \textbf{Type} &  \textbf{Description} & \textbf{Captured Behavior} \\ \cmidrule(lr){1-3}
    Boolean & Usage of installation hook(s) & Arbitrary code execution \\
    Continuous & Number of words in installation scripts & Structural feature of source code\\
    Continuous & Number of lines in installation scripts & Structural feature of source code\\
    Continuous & Number of words in source code files & Structural feature of source code\\
    Continuous & Number of lines in source code files & Structural feature of source code\\
    Continuous & Number of URLs & Security-sensitive string(s)\\
    Continuous & Number of IP addresses & Security-sensitive string(s)\\
    Continuous & Number of suspicious tokens in strings & Security-sensitive string(s)\\
    Continuous & Number of base64 strings & Presence of obfuscation\\
    Continuous & Mean, std. deviation, 3rd quartile, and max value of Shannon entropy of strings in all source code files & Presence of obfuscation\\
    Continuous & Number of homogeneous and heterogenous strings in all source code files & Presence of obfuscation\\
    Continuous & Mean, std. deviation, 3rd quartile, and max value of Shannon entropy of identifiers in all source code files & Presence of obfuscation\\
    Continuous & Number of homogeneous and heterogenous identifiers in all source code files & Presence of obfuscation\\
    Continuous & Mean, std. deviation, 3rd quartile, and max value of Shannon entropy of strings in installation script & Presence of obfuscation\\
    Continuous & Mean, std. deviation, 3rd quartile, and max value of Shannon entropy of identifiers in installation script & Presence of obfuscation\\
    Continuous & Mean, std. deviation, 3rd quartile, and max value of ratio of square brackets per source code file size & String manipulation  \\
    Continuous & Mean, std. deviation, 3rd quartile, and max value of ratio of equal signs per source code file size & String manipulation\\
    Continuous & Mean, std. deviation, 3rd quartile, and max value of ratio of plus signs per source code file size & String manipulation\\
    Continuous & No. of files per selected extensions (91 in total) & Structural feature of the package \\


\end{tabular}
  \caption{Set of features considered for the classification of malicious packages.}\label{tab:feature-set}
  \end{table*}

To identify the features of interest we inspect the malicious samples available in \ac{BKC}~\cite{dasfreakBackstabbersKnife}.
Since our goal is to detect the presence of malicious code in more than one programming language, we do not consider features specifically crafted for a certain programming language (e.g., usage of certain APIs), but we focus on lexical and structural aspects. Thus, we adopt the features proposed in ~\cite{ohm2022feasibility, sejfia2022practical} that have general applicability and are not specific to JavaScript (cf. Table~\ref{tab:relwork-features}). Then, we propose additional features based on expert knowledge.

The final set of features considered in our work is described in Table~\ref{tab:feature-set}. Such features capture characteristics of strings, identifiers, file extensions, and the usage of installation hooks. Their independence from a specific language has the advantage that they do not need constant maintenance as it would be required for language-specific features, like a list of security-related APIs.

We utilize the Pygments lexer\footnote{https://pygments.org/} to parse and process the source code files (\textit{.js, .py}) as well as the installation scripts (\textit{package.json, setup.py}) of the package being analyzed. From such files, we extract the following types of lexical tokens: strings, identifiers, operators, and punctuation. 


In the following, we describe and motivate the choice of the different features.

\textbf{Usage of installation hooks.} As described in Section~\ref{sec:background}, the installation script is commonly used by attackers to trigger execution via installation hooks.
Thus, we detect the presence of the \textit{setup.py} file (for Python) and of the tokens \textit{install, post-install, pre-install} in the \textit{package.json} file (for JavaScript) using a boolean feature.  


\textbf{Code obfuscation.} A property that often characterizes malicious strings is obfuscation. To detect obfuscation and encodings (e.g., base64) in both strings and identifiers, we leverage the Shannon entropy~\cite{10.1145/3560835.3564548} combined with the concept of \ac{GL}~\cite{10.1145/3183713.3196889}. 
Generalization Languages are used in the context of error detection and consist of encoding data using a pre-defined alphabet to abstract specific values into a pattern.
The observation made by Huang et al.~\cite{10.1145/3183713.3196889} is that abstracting specific values into patterns overcomes data sparsity and makes co-occurrence of values more reliable to quantify compatibility (in our case the Shannon entropy) between patterns. 
 
In our case, given a string or identifier (i.e., variable, function, and class names) as input, we transform every character $x$ using the following mapping:
\begin{equation}
  \hbox{$GL_4$}(x)= \begin{cases}
    \hbox{L if } x \hbox{ is a lowercase character} \\
    \hbox{U if } x \hbox{ is an uppercase character} \\
    \hbox{D if } x \hbox{ is a digit} \\
    \hbox{S if } x \hbox{ is a symbol} \\
  \end{cases} 
  \label{eq:GL}
\end{equation}
where $GL_4$ indicates a \ac{GL} with an alphabet of four symbols. 
For example, the string \textit{YmFzaA==} (base64 encoding of \textit{bash}) becomes \textit{ULULLUSS}, whereas a non-encoded string like \textit{while} becomes \textit{LLLLL}. 
%
Once the strings or the identifiers are transformed using $GL_4$, we compute the mean, standard deviation, 3rd quartile, and maximum value of Shannon entropy in both the cases of source code files and installation scripts.
Also, for both strings and identifiers, we count those that are homogeneous (i.e., having all characters equal after $GL_4$ encoding) and heterogeneous (i.e., having at least one different character after $GL_4$ encoding).


\textbf{Sensitive Strings.} 
Malicious code generally introduces strings with certain properties (e.g., URLs, shell commands).
To detect the presence of security-related strings, we adopt the approach presented in~\cite{10.1145/3560835.3564548}. Thus, we build a dictionary of keywords from offensive security cheat sheets (e.g., reverse shells, the path to sensitive files). The keywords are included both in plain text and in different encodings (e.g., base64, base32, rot-13, URL-encoding). The corresponding feature consists of the count of hits to this dictionary.


\textbf{Structural features of source code files and string manipulation.} We capture the code size by counting the number of words and the number of lines of code.

For symbols commonly used to manipulate strings (i.e., plus sign, equal sign, and square brackets) we compute their ratio over the size of the file containing them and we report mean, standard deviation, 3rd quartile, and maximum value for all the files contained in the analyzed package.


\textbf{Structural features of the package.} Since malicious packages may execute malware from external resources to the running script (e.g., binaries, shell scripts), we build a custom list of 91 popular file extensions based on the ones appearing in the malicious packages contained in \ac{BKC}. We report the count of files contained in the analyzed package for these extensions.

\begin{figure}
    \centering
    \includegraphics[scale=0.28]{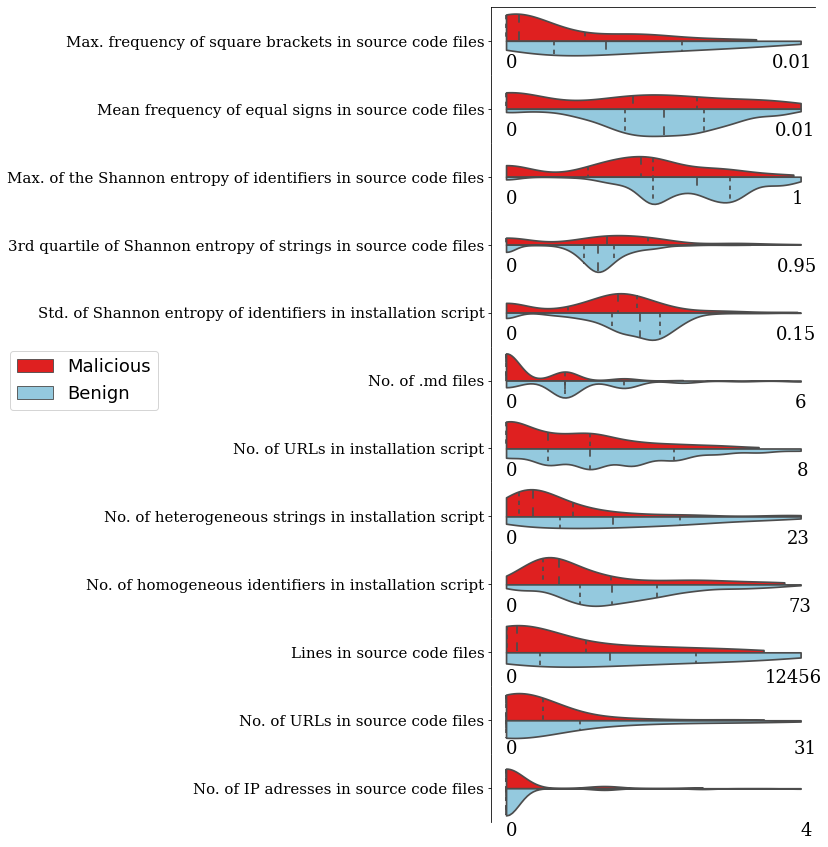}
    \caption{Distribution of a subset of continuous features extracted from benign and malicious samples (both Python and JavaScript). 
    }
    \label{fig:skewed distribution}
  \end{figure}

To support the choice of the features described above, we inspect their statistical distribution.
In particular, we compare the distribution of the features when they are extracted from benign packages and malicious packages. This allows us to initially assess whether the proposed features are able to discriminate between the two classes.
Figure~\ref{fig:skewed distribution} depicts the violin plots for some representative cases. Comparing the distributions in benign and malicious cases, we observe significant differences for most of the features. In the cases where the distributions are similar (e.g., No. of IP addresses in source code files), we still detect that at least one of the statistical measures (e.g., mean, median, minimum value) differs between the malicious and benign cases. 



\begin{figure}
  \centering
  \includegraphics[width=.38\textwidth]{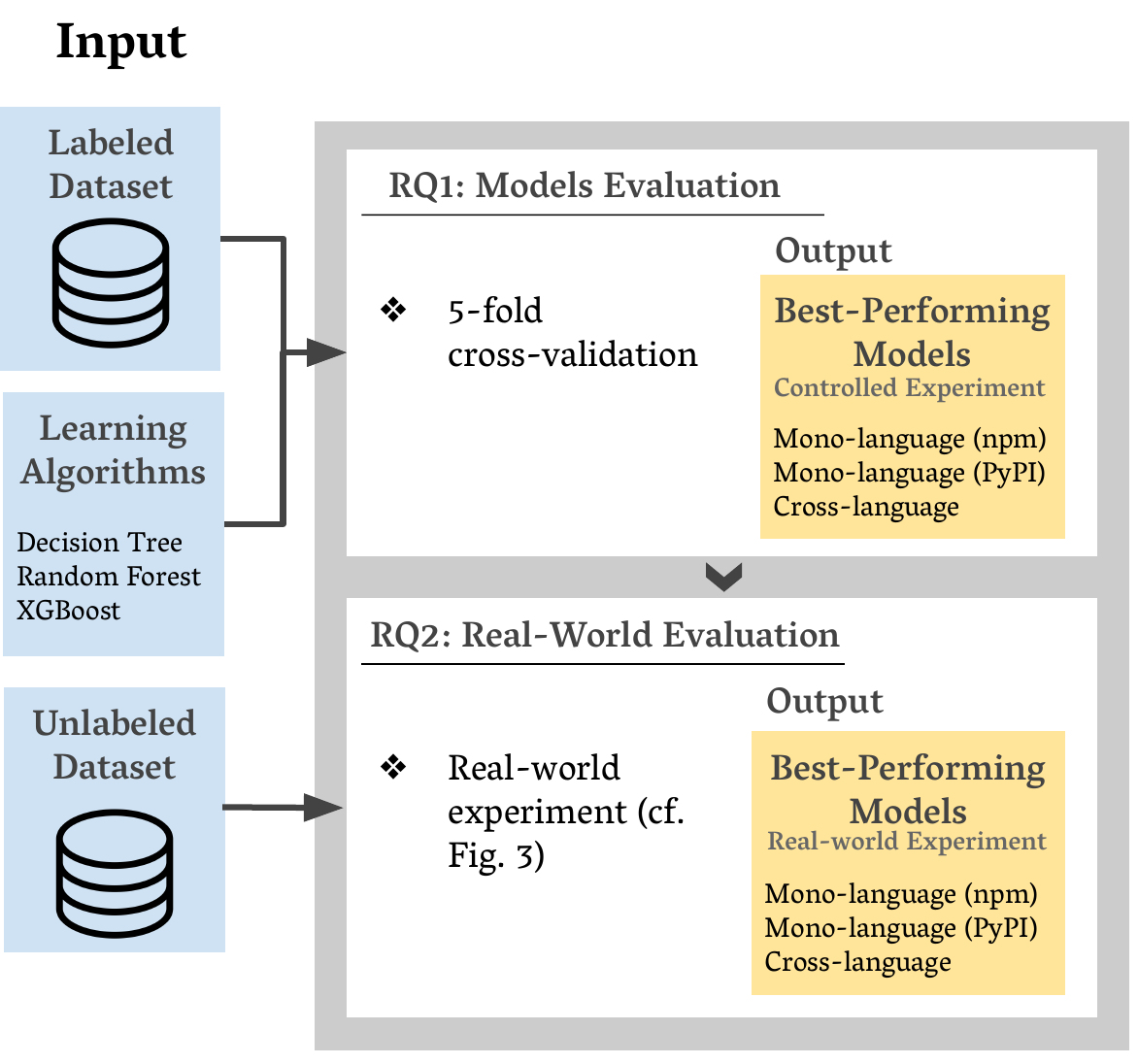}
  \caption{Workflow followed for the evaluation of our approach for the detection of potentially malicious packages.
  }
  \label{fig:approach}
\end{figure}
  

\section{Evaluation} \label{sec:evaluation}



This section answers to the research questions. Figure~\ref{fig:approach} describes the approach followed. Similarly to~\cite{sejfia2022practical,ohm2022feasibility}, we conduct both a controlled and a real-world experiment.

We answer RQ1 by conducting a controlled experiment, in which we use the labeled datasets constructed as described in Section~\ref{sec:datasets} to evaluate the performances of the trained classifiers in both the cases of npm and PyPI. 
Once identified the best-performing models, we evaluated the mono-language and cross-language classifiers by analyzing newly published packages in npm and PyPI, thereby answering RQ2.   

The labeled dataset used in the controlled experiment, the best-performing models from RQ1, and the list of malicious packages found in the wild are available online\footnote{\url{https://github.com/SAP-samples/cross-language-detection-artifacts}}.

\begin{table*}[!hbtp]

  \centering
  \begin{tabular}{rccccccccccccc|}
 
    & \multicolumn{8}{c}{\textbf{JavaScript}} \\
   & \multicolumn{4}{c}{\textbf{Mono-language}} & \multicolumn{4}{c}{\textbf{Cross-language}}\\
   & \multicolumn{4}{c}{(Train set: JS; Test Set: JS)} & \multicolumn{4}{c}{(Train set: JS+PY; Test Set: JS)}\\
  
  & Pr. & Rec. & F1 & Acc.  & Pr. & Rec. & F1 & Acc. \\ \cmidrule(lr){2-5} \cmidrule(lr){6-9} 
 
  DT  & \textbf{100.0±0.0} & 68.0±8.89 & 80.6±6.5 & 96.9±0.9 & 95.9±6.9 & 49.5±17.6 &  63.0±20.1 & 94.8±1.7 \\ 
  RF  & 98.5±3.1  & 53.5±14.7 & 68.1±12.8 & 95.3±1.4 & \textbf{98.5±3.1} & 50.0±16.8 & 64.55±16.1 & 95.0±1.6 \\ 
 \textbf{XGB} & 96.5±4.0  & \textbf{75.5±6.9} & \textbf{84.4±4.2} & \textbf{97.3±0.6} & 97.1±3.8 & \textbf{63.5±10.3} & \textbf{76.3±7.9} & \textbf{96.2±1.0} \\ 

 \\

 & \multicolumn{8}{c}{\textbf{Python}} \\
 & \multicolumn{4}{c}{\textbf{Mono-language}} & \multicolumn{4}{c}{\textbf{Cross-language}}\\
 & \multicolumn{4}{c}{(Train set: PY; Test Set: PY)} & \multicolumn{4}{c}{(Train set: JS+PY; Test Set: PY)}\\
    
 & Pr. & Rec. & F1 & Acc.  & Pr. & Rec. & F1 & Acc. \\ \cmidrule(lr){2-5} \cmidrule(lr){6-9} 

 DT  & 81.6±18.3 & 28.9±14.8 &  39.4±11.2 & 92.0±0.6 & 97.2±8.3 & 16.7±9.6 &  27.4±13.8 & 91.6±1.0 \\ 
 RF  & 79.2±9.4 & 51.7±14.4 & 61.0±9.9 & 93.8±1.0 & 92.5±16.9 & 15.6±8.6 & 25.9±12.9 & 91.6±0.9 \\ 
\textbf{XGB} & \textbf{74.4±13.0}   & \textbf{63.9±13.7} & \textbf{68.0±11.6} & \textbf{94.2±2.0} & \textbf{87.1±11.1}   & \textbf{55.6±13.4} & \textbf{66.9±11.1} & \textbf{94.8±1.5} \\ 

 \end{tabular}
  \caption{Results of 5-fold cross validation experiment for both the cases of JavaScript (JS) and Python (PY). For both mono-language and cross-language models, we report precision (Pr.), recall (Rec.), F1-score (F1), and accuracy (Acc.) in percentages. \label{tab:model-eval-npm}}
  \end{table*}

  

    
   
  

\subsection{RQ1: Models}\label{sec:modelselection}

We evaluate the performances of the classification algorithms described in Section~\ref{sec:meth-classification} to solve the task of detecting potentially malicious packages with a controlled experiment, where labels of data are known.

To address the limitation of small dataset sizes, we follow an approach similar to~\cite{sejfia2022practical} and refrain from keeping a separate hold-out set. Instead, we conduct a 5-fold cross-validation experiment, repeating it ten times. This approach allows us to assess the performance of the models and report the corresponding score metrics.
During each split of the cross-validation, we employ stratified sampling to ensure that the 90-10 ratio between benign and malicious samples is maintained. 


Each learning algorithm (i.e., \ac{DT}, \ac{RF}, \ac{XGBoost}) is trained respectively on the two mono-language datasets and on the cross-language dataset (cf. Section~\ref{sec:datasets}). We obtain a total of 9 classifiers. Table~\ref{tab:model-eval-npm} reports the percentage values - computed on the positive class (i.e., malicious packages) - of precision, recall, F1-score, and accuracy after the 5-fold cross-validation for each ecosystem.

\textbf{Detection in JavaScript.} Considering the mono-language case (i.e., both train and test sets composed of only JavaScript samples), the model providing the highest precision is the one obtained with the \ac{DT} algorithm (i.e., 100\%). However, such a model is subject to a high number of false negatives (recall of 68.0\%) when compared to \ac{XGBoost} (recall of 75.5\%). In fact, the latter model achieves the best trade-off between accuracy and recall (i.e., F1 score of 84.4\%).

For the cross-language case (i.e., train set composed of JavaScript and Python samples, test set composed of JavaScript samples only), the model achieving the highest precision is \ac{RF}. However, the best trade-off is offered also in this case by \ac{XGBoost}.


\textbf{Detection in Python.} In both the case of mono-language models for Python (i.e., train and test sets composed of only Python samples) and cross-language models (i.e., train set composed of JavaScript and Python samples, test set composed of Python samples only), the model with highest precision is \ac{DT}. However, the recall is too low to have practical utility.
Thus, also in this case \ac{XGBoost} outperforms the other models, having a precision of 74.4\%, recall of 63.9\%, and F1-score of 68.0\%.

\begin{Sbox}
\begin{minipage}{0.9\linewidth}
\textbf{Response to RQ1}: As shown in Table~\ref{tab:model-eval-npm}, the XGBoost models exhibit the most favorable balance between precision and recall in both the mono-language and cross-language scenarios for both JavaScript and Python packages.
\end{minipage}
\end{Sbox}

\begin{center}
\fbox{\TheSbox}
\end{center}

By evaluating the performances of the \ac{XGBoost} models both in the mono-language and cross-language case, it appears that the mono-language model performs slightly better than the cross-language one in the classification of JavaScript packages. Vice versa, in the case of Python, the cross-language model based on \ac{XGBoost} performs slightly better than the mono-language one. Since a hold-out set was not used, the comparison between the cross-language and mono-language models, which exhibit the best performance, is addressed in the real-world experiment in Section~\ref{sec:experiment}.

\subsection{RQ2: Real-World Evaluation}\label{sec:experiment}

\begin{figure}
  \centering
 
  \includegraphics[width=.45\textwidth]{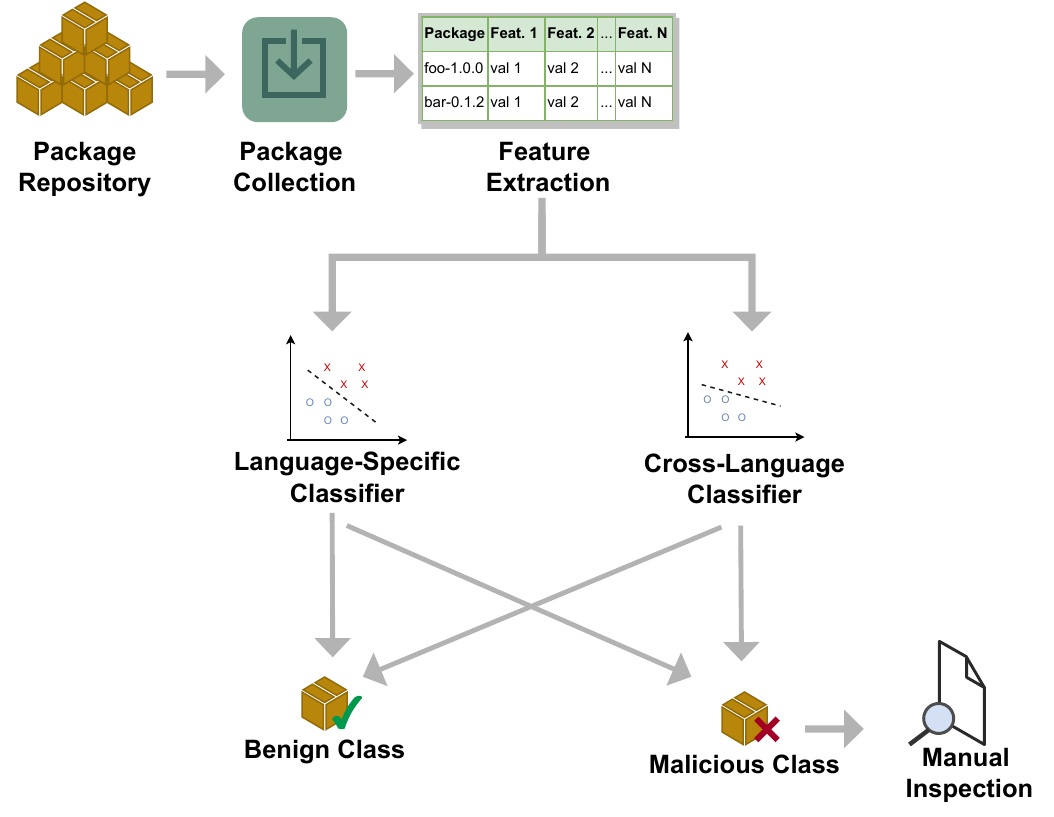}
  \caption{Experiment consisting of the classification of daily-uploaded packages in PyPI and npm for the detection of malicious packages in the wild. The depicted procedure applies to both npm and PyPI.
   }
  \label{fig:maldetection}
\end{figure}

\begin{table*}[!hbtp]

  \centering
  \begin{tabular}{rccccccccccccc|}

   & \multicolumn{4}{c}{\textbf{npm (JavaScript)}} & \multicolumn{4}{c}{\textbf{PyPI (Python)}} \\
  
  & Benign & Malicious (FP) & Malicious (TP) & Pr. & Benign & Malicious (FP) & Malicious (TP) & Pr.\\ \cmidrule(lr){2-5} \cmidrule(lr){6-9} 
 
  Mono-language & 10428 & 1229 & 38 & 3.1\% & 19097 & 485 & 15 & 3.1\% \\
  \textbf{Cross-language} & 10519 & 1083 & 37 & \textbf{3.4\%} & 19196 & 385  & 17 & \textbf{4.4\%}

\end{tabular}
  \caption{Comparison of mono-language and cross-language models in classifying daily-uploaded packages from the wild for 10 days. The best results are highlighted in bold.\label{tab:langdep-vs-langindep}}
  \end{table*}

  \begin{figure*}[!htbp]

    \begin{tabular}{C{.5\textwidth}C{.5\textwidth}}
      
      \begin{large}\textbf{npm}\end{large} & \begin{large}\textbf{PyPI}\end{large}\\
    \parbox{0.48\textwidth}{
              \begin{tikzpicture}
            \def\radius{1cm}
            \def\mycolorbox#1{\textcolor{#1}{\rule{2ex}{2ex}}}
            \colorlet{colori}{blue!100}
            \colorlet{colorii}{green!100}
            
            \coordinate (ceni);
            \coordinate[xshift=\radius] (cenii);
            
            \draw[fill=colori,fill opacity=0.5] (ceni) circle (\radius);
            \draw[fill=colorii,fill opacity=0.5] (cenii) circle (\radius);
            
            
            \node[yshift=10pt] at (current bounding box.north) {Malicious Flags (FP)};
            
            \node at ([xshift=2.5cm] current bounding box.east) 
            {
              
            \begin{tabular}{@{}ll@{\,\,}c@{}}
              & Total: &  \\
            \mycolorbox{colori!50} & Mono-language (npm) =& 1229 \\
            \mycolorbox{colorii!50} & Cross-language =& 1083 \\
            \end{tabular}
            };
            
            \node[xshift=-.5\radius] at (ceni) {$419$};
            \node[xshift=.5\radius] at (cenii) {$275$};
            \node[xshift=.4\radius] at (ceni) {$798$};

            \end{tikzpicture}
            \subcaption{}\label{fig:npm-FP}
    } & 
    \parbox{0.48\textwidth} {
              \begin{tikzpicture}
            \def\radius{1cm}
            \def\mycolorbox#1{\textcolor{#1}{\rule{2ex}{2ex}}}
            \colorlet{colori}{blue!100}
            \colorlet{colorii}{green!100}
            
            \coordinate (ceni);
            \coordinate[xshift=\radius] (cenii);
            
            \draw[fill=colori,fill opacity=0.5] (ceni) circle (\radius);
            \draw[fill=colorii,fill opacity=0.5] (cenii) circle (\radius);
            
            
            \node[yshift=10pt] at (current bounding box.north) {Malicious Flags (FP)};
            
            \node at ([xshift=2.5cm] current bounding box.east) 
            {
              
            \begin{tabular}{@{}ll@{\,\,}c@{}}
              & Total: &  \\
            \mycolorbox{colori!50} & Mono-language (PyPI) =& 561 \\
            \mycolorbox{colorii!50} & Cross-language =& 453 \\
            \end{tabular}
            };
            
            \node[xshift=-.5\radius] at (ceni) {$228$};
            \node[xshift=.5\radius] at (cenii) {$132$};
            \node[xshift=.4\radius] at (ceni) {$251$};

            \end{tikzpicture}
       \subcaption{}\label{fig:PYPI-FP}
    } \\
    
    \parbox{0.48\textwidth} {
              \begin{tikzpicture}
            \def\radius{1cm}
            \def\mycolorbox#1{\textcolor{#1}{\rule{2ex}{2ex}}}
            \colorlet{colori}{blue!100}
            \colorlet{colorii}{green!100}
            
            \coordinate (ceni);
            \coordinate[xshift=\radius] (cenii);
            
            \draw[fill=colori,fill opacity=0.5] (ceni) circle (\radius);
            \draw[fill=colorii,fill opacity=0.5] (cenii) circle (\radius);
            
            
            \node[yshift=10pt] at (current bounding box.north) {Malicious Flags (TP)};
            
            \node at ([xshift=2.5cm] current bounding box.east) 
            {
              
            \begin{tabular}{@{}ll@{\,\,}c@{}}
              & Total: &  \\
            \mycolorbox{colori!50} & Mono-language (npm) =& 38 \\
            \mycolorbox{colorii!50} & Cross-language =& 37 \\
            \end{tabular}
            };
            
            \node[xshift=-.5\radius] at (ceni) {$1$};
            \node[xshift=.5\radius] at (cenii) {$0$};
            \node[xshift=.4\radius] at (ceni) {$37$};

            \end{tikzpicture}
       \subcaption{}\label{fig:npm-TP}
    } & 
    
    \parbox{0.48\textwidth} {
       \begin{tikzpicture}
                \def\radius{1cm}
                \def\mycolorbox#1{\textcolor{#1}{\rule{2ex}{2ex}}}
                \colorlet{colori}{blue!100}
                \colorlet{colorii}{green!100}
                
                \coordinate (ceni);
                \coordinate[xshift=\radius] (cenii);
                
                \draw[fill=colori,fill opacity=0.5] (ceni) circle (\radius);
                \draw[fill=colorii,fill opacity=0.5] (cenii) circle (\radius);
                
                
                \node[yshift=10pt] at (current bounding box.north) {Malicious Flags (TP)};
                
                \node at ([xshift=2.5cm] current bounding box.east) 
                {
                  
                \begin{tabular}{@{}ll@{\,\,}c@{}}
                  & Total: &  \\
                \mycolorbox{colori!50} & Mono-language (PyPI) =& 15 \\
                \mycolorbox{colorii!50} & Cross-language =& 17 \\
                \end{tabular}
                };
                
                \node[xshift=-.5\radius] at (ceni) {$3$};
                \node[xshift=.5\radius] at (cenii) {$5$};
                \node[xshift=.4\radius] at (ceni) {$12$};

                \end{tikzpicture}
        \subcaption{}\label{fig:PYPI-TP}
    } \\
    \end{tabular}
    \caption{Comparison of results for the malicious samples classified by the mono-language and cross-language models in the case of npm. In (a) and (b) the false positives for npm and PyPI respectively. In (c) and (d) the true positives for npm and PyPI respectively. }
    \end{figure*}
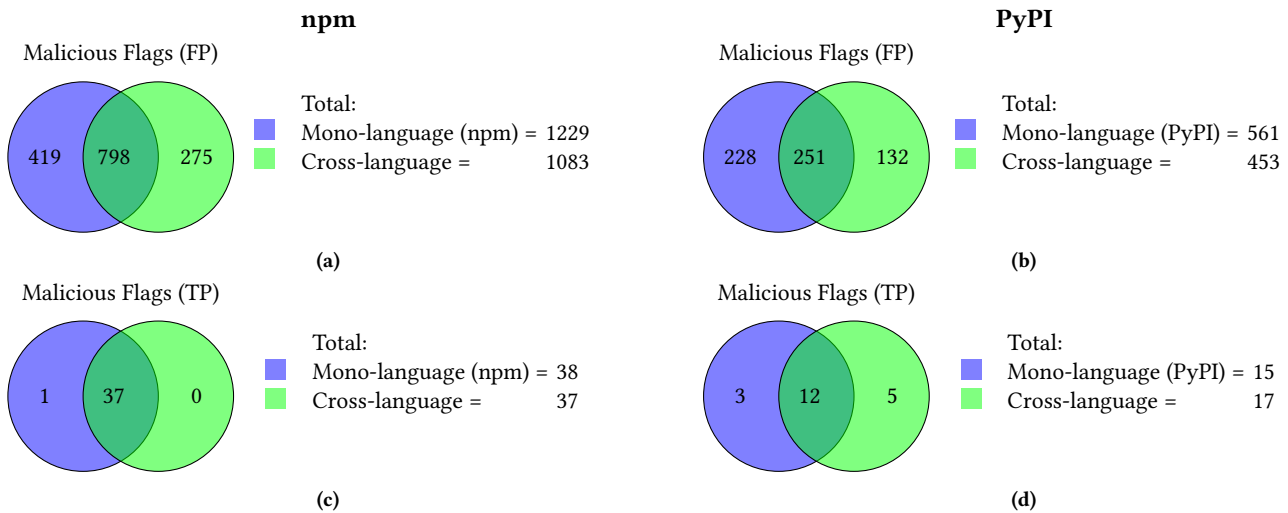     


In this section, we evaluate how the mono-language models and the cross-language model identified in RQ1 (cf. Section~\ref{sec:modelselection}) perform in the classification of benign and malicious packages in the wild.  To do so we conduct the experiment depicted in Figure~\ref{fig:maldetection}. 

We collect newly uploaded packages to npm and PyPI within a period of 10 days (i.e., from Oct. 24 to Nov. 2, 2022). To get the list of packages uploaded to PyPI on a given day we rely on the XML-RPC APIs of the official warehouse\footnote{https://warehouse.pypa.io/api-reference/xml-rpc.html}. Since a comparable feature is not available for npm, we use the RSS feed about new npm packages from \textit{libraries.io}\footnote{https://libraries.io}. For both npm and PyPI, we directly download the packages from the official package repositories.

We extract the features from the downloaded packages as described in Section~\ref{sec:feature-set} and classify such packages using both the mono-language model (of the related ecosystem) and the cross-language model identified in Section~\ref{sec:modelselection}.
For the packages classified as malicious by at least one model, we conduct a manual review to verify the true positives and false positives. In the case of npm packages, we manually inspect \textit{.js}, \textit{.sh}, and \textit{package.json} files while for PyPI packages we inspect \textit{.py} and \textit{.sh} files. In the manual analysis, we look for signs of malicious behavior (e.g., implementation of reverse shells, data exfiltration), and for the obfuscated scripts we attempt to de-obfuscate them.
Table~\ref{tab:langdep-vs-langindep} reports the results of the classification performed by both the mono-language and cross-language models.

In the case of npm, the mono-language model has 419 unique false positives w.r.t. the cross-language model (cf. Figure~\ref{fig:npm-FP}).
The number of unique false positives found by the cross-language model is 275. In terms of true positives, both models correctly classify the same 37 packages as malicious (cf. Figure~\ref{fig:npm-TP}). However, the mono-language model identifies an additional malicious package.  In terms of precision, the cross-language model shows slightly better performance compared to the mono-language model. Overall, we manually reviewed 1,530 packages over 10 days (i.e., 153 packages/day). 


For PyPI, the mono-language model classifies as malicious 231 more packages than the cross-language model. Of these packages, 228 are false positives (cf. Figure~\ref{fig:PYPI-FP}), and 3 are true positives that the cross-language model missed (cf. Figure~\ref{fig:PYPI-TP}). Also in the case of PyPI, the cross-language model has better precision than the mono-language model (4.4\% over 3.1\%, cf. Table~\ref{tab:langdep-vs-langindep} column "Pr."). Overall, we manually reviewed 631 packages over 10 days (i.e., 63 packages/day). 

Though the cross-language model does not drastically improve the precision compared to the mono-language models, the difference in terms of manual effort is non-negligible as the number of false positives considerably decreases without a major loss in terms of true positives. 

All the malicious packages (true positives) found in npm (38 in total) and PyPI (20 in total) have been reported to the respective security teams following the official channels (cf. Section~\ref{sec:resp-disclosure}). As a result of our reporting, after triaging by the respective npm and PyPI security teams, the 58 reported malicious packages have been removed from the repositories and are no longer available for download. As we believe in the importance of maintaining (for research purposes) a database of malicious packages related to \ac{OSS} supply chain attacks, we contributed to \ac{BKC} by uploading the detected malicious packages. 


\begin{Sbox}
\begin{minipage}{0.9\linewidth}
\textbf{Response to RQ2}: By classifying packages daily uploaded to npm and PyPI within a period of 10 days we found that the cross-language model achieves better precision than the mono-language models. In total, we find 58 previously unknown malicious packages (38 for npm and 20 for PyPI).
\end{minipage}
\end{Sbox}

\begin{center}
\fbox{\TheSbox}
\end{center}

\section{Discussion}\label{sec:discussion}

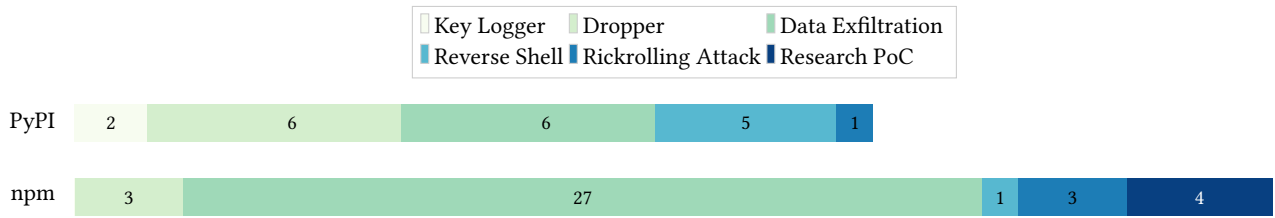
\begin{figure*}
  \centering

\begin{tikzpicture}

  \definecolor{darkgray176}{RGB}{176,176,176}
  \definecolor{gainsboro212238205}{RGB}{212,238,205}
  \definecolor{honeydew247252240}{RGB}{247,252,240}
  \definecolor{lightgray204}{RGB}{204,204,204}
  \definecolor{mediumturquoise87184208}{RGB}{87,184,208}
  \definecolor{midnightblue864129}{RGB}{8,64,129}
  \definecolor{silver159217184}{RGB}{159,217,184}
  \definecolor{steelblue29125182}{RGB}{29,125,182}
  
  \begin{axis}[
  axis line style={draw=none},
  tick style={draw=none},
  legend cell align={left},
  legend columns=3,
  legend style={
    fill opacity=1,
    draw opacity=1,
    text opacity=1,
    at={(0.5,1.05)},
    anchor=south,
    draw=lightgray204
  },
  tick align=outside,
  tick pos=left,
  xmin=0, xmax=33.6,
  y grid style={darkgray176},
  ymin=-0.5, ymax=1.5,
  ytick style={color=black},
  ytick={0,1},
  yticklabels={npm,PyPI},
  xticklabel=\empty,
  width=1\textwidth,height=.2\textwidth
  ]
  \draw[draw=none,fill=honeydew247252240] (axis cs:0,-0.25) rectangle (axis cs:0,0.25);
  \addlegendimage{ybar,ybar legend,draw=none,fill=honeydew247252240}
  \addlegendentry{Key Logger}
  
  \draw[draw=none,fill=honeydew247252240] (axis cs:0,0.75) rectangle (axis cs:2,1.25);
  \draw[draw=none,fill=gainsboro212238205] (axis cs:0,-0.25) rectangle (axis cs:3,0.25);
  \addlegendimage{ybar,ybar legend,draw=none,fill=gainsboro212238205}
  \addlegendentry{Dropper}
  
  \draw[draw=none,fill=gainsboro212238205] (axis cs:2,0.75) rectangle (axis cs:9,1.25);
  \draw[draw=none,fill=silver159217184] (axis cs:3,-0.25) rectangle (axis cs:25,0.25);
  \addlegendimage{ybar,ybar legend,draw=none,fill=silver159217184}
  \addlegendentry{Data Exfiltration}
  
  \draw[draw=none,fill=silver159217184] (axis cs:9,0.75) rectangle (axis cs:16,1.25);
  \draw[draw=none,fill=mediumturquoise87184208] (axis cs:25,-0.25) rectangle (axis cs:26,0.25);
  \addlegendimage{ybar,ybar legend,draw=none,fill=mediumturquoise87184208}
  \addlegendentry{Reverse Shell}
  
  \draw[draw=none,fill=mediumturquoise87184208] (axis cs:16,0.75) rectangle (axis cs:21,1.25);
  \draw[draw=none,fill=steelblue29125182] (axis cs:26,-0.25) rectangle (axis cs:29,0.25);
  \addlegendimage{ybar,ybar legend,draw=none,fill=steelblue29125182}
  \addlegendentry{Rickrolling Attack}
  
  \draw[draw=none,fill=steelblue29125182] (axis cs:21,0.75) rectangle (axis cs:22,1.25);
  \draw[draw=none,fill=midnightblue864129] (axis cs:29,-0.25) rectangle (axis cs:33,0.25);
  \addlegendimage{ybar,ybar legend,draw=none,fill=midnightblue864129}
  \addlegendentry{Research PoC}
  
 
  \draw (axis cs:1,1) node[
    scale=0.864,
    text=black,
    rotate=0.0
  ]{2};
  \draw (axis cs:1.5,0) node[
    scale=0.864,
    text=black,
    rotate=0.0
  ]{3};
  \draw (axis cs:6,1) node[
    scale=0.864,
    text=black,
    rotate=0.0
  ]{6};
  \draw (axis cs:14,0) node[
    scale=0.864,
    text=black,
    rotate=0.0
  ]{27};
  \draw (axis cs:13,1) node[
    scale=0.864,
    text=black,
    rotate=0.0
  ]{6};
  \draw (axis cs:25.5,0) node[
    scale=0.864,
    text=black,
    rotate=0.0
  ]{1};
  \draw (axis cs:18.5,1) node[
    scale=0.864,
    text=black,
    rotate=0.0
  ]{5};
  \draw (axis cs:27.5,0) node[
    scale=0.864,
    text=black,
    rotate=0.0
  ]{3};
  \draw (axis cs:21.5,1) node[
    scale=0.864,
    text=black,
    rotate=0.0
  ]{1};
  \draw (axis cs:31,0) node[
    scale=0.864,
    text=white,
    rotate=0.0
  ]{4};
  \end{axis}
  
  \end{tikzpicture}

  \caption{Types of malicious behavior among the 58 findings between npm and PyPI.}
  \label{fig:malwaretypes}
\end{figure*}

In this section, we describe the characteristics of the malicious packages found in npm and PyPI. We report some remarks on the two ecosystems observed while inspecting the false positives. Finally, we highlight notable differences in terms of features between malicious true positives, false positives, and packages flagged as benign.

\textbf{Malware types.} 
Figure~\ref{fig:malwaretypes} depicts the types of malicious behaviors found in both npm and PyPI during our real-world experiment (cf. Section~\ref{sec:experiment}).

The majority of malicious packages both in npm and PyPI aim at achieving data exfiltration (27 and 6 packages respectively) and the most common exfiltrated information consists of public IP address and environment variables.
Other common malicious behaviors that we observed involve the opening of a reverse shell and the download and execution of malicious code (i.e., dropper). The latter is commonly achieved by making an HTTP(S) request to download and execute a second-stage payload. In a malicious discovery within npm, we identified a novel tactic (in comparison to those in the \ac{BKC}).
This tactic involves placing the second-stage payload within the TXT field of a DNS response, under the control of the attacker. Subsequently, a request to this DNS is initiated through a third-party service (i.e., Google's DNS resolver). 
In this way, the attacker is less likely to have the request from the victim blocked by any firewall.

In npm, we found also research proofs-of-concept, i.e., the attacker aims at achieving execution to demonstrate the potential risks when installing their package. One notable example is a package containing code obfuscated with AES-256 encryption, which is decrypted at runtime. Upon execution, the code creates a new file on the victim's machine containing a warning message about the potential harm.
Furthermore, we identified 4 packages (3 in npm and 1 in PyPI) conducting a Rickrolling attack at installation time (i.e., starting the reproduction of the song \textit{Never Gonna Give You Up} from Rick Astley). When we reported these, the respective security teams decided not to remove the said packages as they did not consider Rickrolling as violating their term of service.

Although this behavior was not part of the ones available in the training dataset (cf. Section~\ref{sec:datasets}), our approach identified two keyloggers in PyPI.

\textbf{Malware campaigns.} During the analysis of daily uploaded packages to npm and PyPI, we observed the presence of several malware campaigns. Additionally, we consider variations of the malicious code that only differ in the subdomain of the URLs.

In the case of npm, we detected 7 campaigns of which the largest includes 12 packages. Only in one campaign the attacker experiment with different ways to achieve code execution, while in all the other cases both the malicious code and the way of triggering its execution are the same. 
In the case of PyPI, we detected 5 campaigns of which the largest includes 4 packages.

It is worth mentioning that we have detected a case of a cross-language campaign, meaning that the same malicious behavior was present in both npm and PyPI. This finding suggests that attackers may spread their malicious activities across multiple ecosystems in order to increase their chances of success.

\textbf{Malicious code obfuscation.} The majority of malicious packages found both in npm and PyPI during our experiment do not obfuscate the malicious code.
Among the 38 malicious packages from npm, only 2 obfuscate the code using AES encryption, and 2 packages use custom obfuscation (e.g., renaming sensitive functions like exec in human-unreadable ways).
Among the 20 malicious packages found in PyPI, 3 use simple obfuscation techniques (e.g., encoding the source code in bytes, base64, rot13) and 3 other packages employ custom obfuscation.

\textbf{Analysis of false positives.}
The majority of false positives in both npm and PyPI consist of small packages (often containing only \textit{package.json} or \textit{setup.py}) with almost no functionality (e.g. print of "hello world" or sum of numbers). Possible use cases of such packages are to reserve names (e.g., future projects, prevent typosquatting) in the package repository or to test the upload of packages.

We also observe the concept of campaign (i.e., packages with the same behavior) for some non-malicious packages. A curious case observed in npm is about a set of packages whose sole purpose is to use the \textit{give-me-a-joke} project as a dependency, probably with the aim of increasing its popularity. Another interesting finding is about a package containing nothing but the CV of its creator.

In 4 packages (3 in npm and 1 in PyPI) we detect the presence of obfuscated code.
Since we could not observe obvious malicious behaviors we flagged these as false positives.

Finally, we acknowledge that the number of false positives (cf. Table~\ref{tab:langdep-vs-langindep}) clearly needs further improvement, although they are still manageable for a manual review, especially considering that the majority of them were small in size.
 

\textbf{Comparison on features.} 
While $81\%$ of the malicious packages found in npm (true positives) make use of installation hooks, only $8\%$ and $2\%$ make use of them among the false positives and the packages flagged as benign, respectively.
Also in the case of PyPI, the majority of true positives (i.e., $58\%$) make use of installation hooks, while it is less common in the case of false positives and packages flagged as benign (i.e., $5\%$ and $1\%$, respectively).
Another aspect concerns the presence of Markdown files: only a minority of true positives (i.e., $25\%$ for npm and $42\%$ for PyPI) contain at least one \textit{.md} file, whereas most of the packages flagged as benign contain such files ($82\%$ for npm and $79\%$ for PyPI).

\textbf{Cross-language vs. Mono-language models.} As observed in the real-world evaluation (cf. Section~\ref{sec:experiment}), the cross-language model exhibits greater precision compared to the mono-language models. To delve into the rationale behind this, we explored the relationship between feature importance across models and the traits of the malicious packages identified by each model.

The usage of installation hooks and the count of Markdown files rank within the most influential features for the mono-language model of npm, whereas they hold less significance for the mono-language model of PyPI. When merging npm and PyPI samples to train the cross-language model, the importance of these features reemerges.
Considering that a substantial portion of true positives in PyPI use installation hooks or lack Markdown files, this characteristic likely contributes to the improvement of the cross-language model's performance compared to the mono-language model specifically designed for PyPI.

Features associated with the size of source code files (e.g., \textit{number of lines}) and the count of homogeneous strings in source code files (cf. Section~\ref{sec:feature-set}) rank within the most influential features for the mono-language model of PyPI, whereas they hold less significance for the mono-language model of npm. Merging npm and PyPI samples to train the cross-language model, the importance of these features reemerges. 
Given that several false positives in npm lack script files and consist solely of the \textit{package.json}, this characteristic is likely a contributing factor in augmenting the cross-language model's precision compared to the mono-language model for npm.

\textbf{Porting to other languages.} In our work, we explored the feasibility of detecting malicious packages across multiple ecosystems. As a preliminary analysis, we focused on the case of npm and PyPI. To achieve this, we defined a set of features that are not specific to npm or PyPI, making it feasible to port our approach to other ecosystems. The only feature required during porting is that the target ecosystem supports the use of install hooks, which is also one of the most important features for the trained models. Examples of ecosystems that support this functionality include Composer for PHP~\cite{composerScripts} and Gem for Ruby (by extending the definition of 'usage of installation hook(s)' to include the usage of extensions by gems~\cite{rubygemsGemsWith}).

\section{Responsible Disclosure}\label{sec:resp-disclosure}
npm and PyPI have two different ways of officially reporting malicious packages. 

In npm, reporting is done via the UI of the official website: the user has to search for the package to be reported, navigate to the project page, click on the 'Report Malware' button, and finally fill in the report form. This procedure can make it difficult to report multiple malwares at the same time (especially since malware campaigns are common). Moreover, multiple reports at the same time enable CAPTCHA tests. All 38 malicious packages found during our analysis have been responsibly reported to the security team according to the abovementioned procedure. Except for the packages conducting a Rickrolling attack (whose behavior was not considered a violation of npm's terms of use), they have been removed from the package repository.   

In PyPI, the official procedure to report malicious packages consists of sending an e-mail to the security team\footnote{\url{https://pypi.org/security/}} with the names of the packages and (preferably) the link to the lines of code containing the malware highlighted using \textit{Inspector}\footnote{\url{https://inspector.pypi.io}}. Also in this case, we responsibly reported the 20 malicious packages found during our analysis. Following the triaging procedure, the PyPI security team confirmed our findings and removed them from the package repository.

\section{Threats to Validity}\label{sec:threats-validity}

Though the results obtained in evaluating our cross-language approach are promising, there are some threats to validity worth highlighting.

Both the mono-language and cross-language models have been trained on previously known malicious samples from \ac{BKC}~\cite{dasfreakBackstabbersKnife}. This has introduced a bias on the type of malware identifiable through our classification (cf. Section~\ref{sec:datasets}), and a limitation of the approach is the difficulty of finding new (unseen) types of malware. Furthermore, attackers could potentially evade detection by carefully constructing malicious packages so that the values of their features (cf. Section~\ref{sec:feature-set}) mimic the distribution seen in benign packages.

In the real-world experiment, we do not manually review (due to their large number) the packages flagged as benign by our classifiers to check the presence of false negatives so we cannot draw conclusions in terms of recall. Nevertheless, during our analysis, we learned from the news~\cite{phylumPhylumDiscovers} about the discovery of 12 malicious packages uploaded to PyPI in the same period of our experiment. By inspecting these cases we observed that none of them were flagged as malicious by our classifiers (both mono-language and cross-language). The peculiarity of these missed packages - compared to those known from \ac{BKC} - is that they are clones of benign packages (thus containing largely benign code) into which a line of malicious code has been injected. 

As described in Section~\ref{sec:discussion}, in the manual analysis of packages collected in the wild and flagged as malicious, we encountered some suspicious packages heavily obfuscated, that we marked as false positives as we were not able to identify malicious behavior. However, a deeper analysis could reveal malicious intent and thus we could have misattributed some false positives.

As mentioned in Section~\ref{sec:datasets}, since there is no ground truth for benign packages in npm and PyPI, we made some assumptions and did our best effort to check that the benign samples in our dataset do not include malicious code. However, we cannot exclude that some packages considered benign may be hiding malicious code and thus the corresponding label in our dataset would be wrong. 

Related to the imbalance problem, there is no certainty about the actual ratio of benevolent to malevolent packages in package repositories. In our work, we assume a ratio of 90\% benign samples and 10\% malicious ones as suggested in~\cite{shabtai2009detection, moskovitch2008unknown}, but other options have been proposed, e.g., 1\% of malicious packages in \cite{ohm2022feasibility}. As we did not conduct experiments on changing such ratios, we cannot evaluate the impact of different ratios on our classification. 

In our work, we evaluate mono-language and cross-language models. However, both are trained using language-independent features, thus the results may differ with mono-language models trained on language-specific features. We consider our approach complementary to those using language-specific features: we cope with the scarcity of samples to increase the set of known malicious samples, required for language-specific approaches. 

\section{Related Works}\label{sec:relatedworks}

Our work proposes a supervised learning approach for the detection of potentially malicious packages in software repositories (i.e., npm and PyPI) to counter \ac{OSS} supply chain attacks. A comparison to the closest works is shown in Table~\ref{tab:relwork-features}. We further discuss works that focus on the problem of malicious code in the context of \ac{OSS} supply chain security. 

Sejfia et al.~\cite{sejfia2022practical} propose a supervised learning approach combined with a code reproducer and a simple clone detector for the automated detection of malicious packages in npm. Although they also consider some language-generic features (e.g., presence of minified code, binary files), their main focus is on language-dependent aspects (e.g., use of specific APIs) for JavaScript. 


Ohm et al.~\cite{ohm2022feasibility} conducted an extensive evaluation of 25,210 models to assess the feasibility of utilizing supervised learning techniques for the detection of malicious packages. They specifically focus on the npm ecosystem and consider mainly language-dependent features next to more generic ones. 

Compared to~\cite{sejfia2022practical} and~\cite{ohm2022feasibility}, we only consider language-independent features that can be easily applied both to JavaScript and Python and we consider \ac{DT} and \ac{RF} as they are the best-performing models in~\cite{sejfia2022practical} and~\cite{ohm2022feasibility}, respectively. 
Since the models and datasets are not available, we cannot compare our models with theirs.

\begin{table}[!hbtp]

  \centering
  \begin{tabular}{cccc}

    & \multicolumn{1}{c}{\textbf{Our work}} & \multicolumn{1}{p{1.3cm}}{\textbf{Sejfia et al. \cite{sejfia2022practical}}} & \multicolumn{1}{p{1.3cm}}{\textbf{Ohm et al. \cite{ohm2022feasibility}}} \\
   \cmidrule(lr){2-4}

   \multicolumn{1}{l}{\textbf{Target Language:}}& \multicolumn{1}{p{1.5cm}}{Python, JavaScript}  & JavaScript & JavaScript\\ 
   \cmidrule(lr){1-4} 
   \multicolumn{1}{l}{\textbf{Analysis type:}}& Static  & Static & Static \\
   \cmidrule(lr){1-4} 
   \multicolumn{1}{p{3cm}}{\textbf{Features: }}\\

   \multicolumn{1}{p{3cm}}{Use of installation hooks} & \checkmark & \checkmark & \checkmark \\

   \multicolumn{1}{p{3cm}}{No. of words/lines in installation scripts} & \checkmark &   &  \\
   \multicolumn{1}{p{3cm}}{No. of words/lines in source code} & \checkmark &   &  \\

   \multicolumn{1}{p{3cm}}{Presence of URLs} & \checkmark &  & \checkmark \\

   \multicolumn{1}{p{3cm}}{Presence of IP addresses} & \checkmark &  & \checkmark \\

   \multicolumn{1}{p{3cm}}{Presence of Base64 strings} & \checkmark &  & \checkmark \\

   \multicolumn{1}{p{3cm}}{Presence of sensitive strings} & \checkmark &   & \checkmark \\

   \multicolumn{1}{p{3cm}}{Presence of obfuscation} & \checkmark & \checkmark  &  \\

   \multicolumn{1}{p{3cm}}{Presence of executables} & \checkmark & \checkmark  &  \\

   \multicolumn{1}{p{3cm}}{Count of script files} & \checkmark &   & \checkmark \\

   \multicolumn{1}{p{3cm}}{Count of files per extension} & \checkmark &   &\\ 

   \multicolumn{1}{p{3cm}}{Stats on square brackets} & \checkmark &   & \checkmark \\

   \multicolumn{1}{p{3cm}}{Stats on equal sign} & \checkmark &   &  \\

   \multicolumn{1}{p{3cm}}{Stats on plus sign} & \checkmark &   & \\

   \multicolumn{1}{p{3cm}}{Use of sensitive APIs} &  & \checkmark  & \checkmark \\

\end{tabular}
  \caption{Comparison with related works~\cite{sejfia2022practical, ohm2022feasibility} in the context of \ac{ML} approaches applied to the detection of malicious packages in package repositories.\label{tab:relwork-features}}
  \end{table}

  In another work, Ohm et al.~\cite{DBLP:journals/corr/abs-2011-02235} describe an approach to detect malicious package campaigns by unsupervised signature generation relative to code reuse. In particular, they generate the \ac{AST} from npm packages and cluster them so that they can identify packages sharing commonalities. In our work, we only focus on lexical features, and our approach is based on supervised learning. In addition, since our focus is cross-language, our target language is not limited to JavaScript.

  Still, Ohm et al.~\cite{10.1145/3407023.3409183} describe a dynamic analysis approach for the detection of malicious JavaScript and Python packages through the analysis of forensic artifacts. Compared to their work, our approach is only static and more lightweight.
  
  Garret et al.~\cite{garrett2019detecting} apply an anomaly detection approach to identify malicious updates in the npm ecosystem. Their use case is to compare a version of a JavaScript package with previous ones to detect malicious updates. They only consider JavaScript-specific features related to whether or not code is added. Instead, our purpose is to determine whether a package is malicious as is, regardless of its relationship to previous versions.
  
  Duan et al.~\cite{duan2021measuring} propose a pipeline based on static and dynamic analysis for the detection of malicious packages in interpreted languages (i.e., JavaScript, Python, Ruby). In particular, they derive heuristic rules from a previous study on supply chain attacks. Though they also target malware detection on multiple languages, compared to them we propose a lightweight approach based on machine learning, that does not require updating and maintaining a list of APIs for each supported language, and that does not require code execution.
  
  In the context of Java, Ladisa et al.~\cite{ladisajavamalware} propose several indicators of malicious behavior that can be observed from Java bytecode. Apart from the different scope, we take the idea of applying Shannon entropy to strings to detect the presence of obfuscation and extend it to identifiers. In addition, we construct the list of suspicious tokens taking cues from their approach.

  Fass et al.~\cite{fass2018jast} aim at the detection of obfuscated code in JavaScript through the extraction of features from the \ac{AST} and the training of a \ac{RF} classifier.
  Compared to them, our approach to detect obfuscation (e.g., encoded strings) focuses only on lexical features.
  
  Pfretzschner et al.~\cite{10.1145/3098954.3120928} describe JavaScript language features that can be exploited by malicious dependencies to conduct attacks. Their approach to detection is static and only focuses on JavaScript.

  Vu et al.~\cite{10.1145/3468264.3468592} approach the detection of malicious injections in Python packages by analyzing discrepancies between source code and built version. Scalco et al.~\cite{onfeasibilitynpmvu} transfer this approach in the context of JavaScript. Since obtaining the source code corresponding to a package is not straightforward~\cite {9678526}, our goal is to determine whether a package is malicious without relying on additional, possibly unavailable, inputs.

Still, Vu et al.~\cite{vu2022pythonbench} interview PyPI's administrators to explore the security goals of such package repository and create a benchmark dataset to review the current malware checker tools available for PyPI (e.g., Bandit4Mal\footnote{https://github.com/lyvd/bandit4mal}, OSSGadget\footnote{https://github.com/microsoft/OSSGadget}). They report that package repositories need malware detectors with very low false-positive rates and thus the need for improvements in detection capabilities.

Although our focus is on pre-built packages, 
it is worth mentioning that different works for the detection of \ac{OSS} software supply chain attacks have been proposed also in the context of source code repositories~\cite{9402087,cao2022fork,vu2020towards}.

\section{Conclusion and Future Works}\label{sec:conclusion}

In this work, we investigate the feasibility of a cross-language approach for the detection of malicious packages in npm and PyPI. Our approach is lightweight and focuses on lexical aspects of code (e.g., characteristics of strings) and structural features of packages (e.g., number of files per extension). First, we define a set of simple yet effective features to discriminate between malicious and benign packages in JavaScript and Python.
Then, we analyze the performances of learning algorithms based on \ac{DT} for the classification of malicious and benign packages through a controlled experiment (i.e., where labels are known) and found that \ac{XGBoost} performs better in all the contexts (i.e., using mono-language datasets for JavaScript or Python, and a cross-language dataset). 
Finally, we evaluated the mono-language and cross-language models by classifying packages in the wild. We identified 58 previously unknown malicious packages and reported them to the respective package repositories.


Considering the encouraging outcomes of our assessment of cross-language detection feasibility, our future focus aims to enhance detection accuracy. This involves delving into supplementary machine learning algorithms to find the best-performing model and expanding feature sets to further refine our approach. Furthermore, we envisage combining the \ac{ML}-based classifier with the automatic analysis of packages associated with the same author of confirmed malicious packages. In fact, with a first manual inspection of the packages uploaded by the same author as those detected in our experiment, we were already able to find additional malicious packages (uploaded in a time window prior to our experiment), either related to earlier versions of the same package or other packages from the same campaign.
In addition, we intend to extend our evaluation to include other languages (e.g., Ruby, PHP).


\section{Data Availability}
The labeled dataset used for the training (cf. Section~\ref{sec:datasets}), the best-performing models obtained in RQ1, and the list of malicious packages found in the wild are available online\footnote{\url{https://github.com/SAP-samples/cross-language-detection-artifacts}}. 

\hfill \break
\small\noindent\textbf{Acknowledgements.}
We thank the reviewers for their constructive feedback, which has greatly contributed to the improvement of our work.
This work is partly funded by EU grants No. 952647 (AssureMOSS) and No. 101120393 (Sec4AI4Sec).
\normalsize

\begin{acronym}[TDMA]
    \acro{AV}{Anti Virus}
    \acro{AVs}{Anti Viruses}
    \acro{CTI}{Cyber Threat Intelligence}
    \acro{C2}{Command and Control}
    \acro{CPG}{Code Property Graph}
    \acro{TUF}{The Update Framework}
    \acro{PKI}{Public Key Infrastructure}
    \acro{CI}{Continuous Integration}
    \acro{CD}{Continuous Delivery}
    \acro{UI}{User Interface}
    \acro{VCS}{Versioning Control System}
    \acro{VMs}{Virtual Machines}
    \acro{VA}{Vulnerability Assessment}
    \acro{VCS}{Version Control System}
    \acro{SCM}{Source Control Management}
    \acro{IAM}{Identity Access Management}
    \acro{CDN}{Content Delivery Network}
    \acro{UX}{User eXperience}
    \acro{SLR}{Systematic Literature Review}
    \acro{SE}{Social Engineering}
    \acro{MITM}{Man-In-The-Middle}
    \acro{SBOM}{Software Bill of Materials}
    \acro{MFA}{Multi-Factor Authentication}
    \acro{AST}{Abstract Syntax Tree}
    \acro{RASP}{Runtime Application Self-Protection}
    \acro{OS}{Operating System}
    \acro{OSS}{Open-Source Software}
    \acro{TARA}{Threat Assessment and Remediation Analysis}
    \acro{CAPEC}{Common Attack Pattern Enumeration and Classification}
    \acro{DoS}{Denial of Service}
    \acro{SCA}{Software Composition Analysis}
    \acro{SLSA}{Supply-chain Levels for Software Artifacts}
    \acro{SDLC}{Software Development Life-Cycle}
    \acro{ICT}{Information and Communication Technologies}
    \acro{C-SCRM}{Cyber Supply Chain Risk Management}
    \acro{DDC}{Diverse Double-Compiling}
    \acro{OSINT}{Open Source Intelligence}
    \acro{U/C}{Utility-to-Cost}
    \acro{LOC}{Lines Of Code}
    \acro{JVM}{Java Virtual Machine}
    \acro{LFI}{Local File Inclusion}
    \acro{PII}{Personally Identifying Information}
    \acro{RCE}{Remote Code Execution}
    \acro{JIT}{Just-In-Time}
    \acro{DoS}{Denial of Service}
    \acro{JAR}{Java Archive}
    \acro{AOT}{Ahead-Of-Time}
    \acro{DT}{Decision Tree }
    \acro{RF}{Random Forest }
    \acro{XGBoost}{eXtreme Gradient Boosting}
    \acro{BO}{Bayesian Optimizer}
    \acro{GL}{Generalization Language}
    \acro{GL3}{}
    \acro{GL4}{}
    \acro{ML}{Machine Learning}
    \acro{BKC}{Backstabber's Knife Collection}
    \acro{KS}{Kolmogrov-Smirnov}
\end{acronym}

\bibliographystyle{ACM-Reference-Format}
\bibliography{bibliography}

\end{document}